\documentstyle[12pt,epsfig]{article}                                                           
                                                           
\newlength{\dinwidth}                                                           
\newlength{\dinmargin}                                                           
\def\lapproxeq{\lower .7ex\hbox{$\;\stackrel{\textstyle                                                           
<}{\sim}\;$}}                                                           
\def\gapproxeq{\lower .7ex\hbox{$\;\stackrel{\textstyle                                                           
>}{\sim}\;$}}  
\newcommand{\porpbar}
{\!\,^{\scriptscriptstyle(}$\mbox{$\bar{p}$}$\,^{\scriptscriptstyle)}}                                                                                         
\def\be{\begin{equation}}                                                           
\def\ee{\end{equation}}                                                           
\def\bea{\begin{eqnarray}}                                                           
\def\eea{\end{eqnarray}}

\begin{document}                                                           
\titlepage                                                           
\begin{flushright}                                                           
IPPP/01/09 \\     
DCPT/01/18 \\                                                           
20 February 2001 \\                                                           
\end{flushright}                                                           
                                                           
\vspace*{2cm}                                                           
                                                           
\begin{center} 
\renewcommand{\thefootnote}{\fnsymbol{footnote}}                                                          
{\Large \bf Rates for rapidity gap Higgs production\footnote[2]{Based on a 
talk by V.A. Khoze at the First Workshop on Forward Physics and Luminosity Determination 
at LHC, Helsinki, 31 October -- 4 November, 2000.}} \\  
                                                           
\vspace*{1cm}                                                           
V.A. Khoze$^a$, A.D. Martin$^a$ and M.G. Ryskin$^{a,b}$ \\                                                           
                                                          
\vspace*{0.5cm}                                                           
$^a$ Department of Physics and Institute for Particle Physics Phenomenology, University of     
Durham, Durham, DH1 3LE \\        
$^b$ Petersburg Nuclear Physics Institute, Gatchina, St.~Petersburg, 188300, Russia                   
\end{center}                                                           
                                                           
\vspace*{2cm}                                                           
                                                           
\begin{abstract}                                                           
We present model-independent estimates of the signal-to-background ratio for Higgs $\rightarrow 
b\bar{b}$ detection in double-diffractive events at the Tevatron and the LHC.  For the missing-mass 
approach to be able to identify the Higgs boson, it will be necessary to tag the $b$ quark jets in the 
central region.  The signal is predicted to be very small at the Tevatron, but observable at the LHC.  
We note that the double-diffractive dijet production, may serve as a unique gluon factory.  This process 
can be used also as a Pomeron-Pomeron luminosity monitor.
\end{abstract}                                                 
            
\newpage                  
\section{Introduction}

It looks quite appealing to study central production processes with a large rapidity gap on either side in 
high energy hadron collisions.  These embrace double-diffractive reactions generated by \lq 
Pomeron-Pomeron\rq\ fusion, searches for new objects (such as the Higgs boson), and new approaches to study 
conventional physics, including the investigation of 
subtle aspects of QCD.  Equally interesting are other processes mediated by the 
colourless $t$-channel exchanges, especially the $W$-boson fusion reactions.

One possibility, which at first sight looks attractive, is the search for the Higgs boson in 
double rapidity gap events at proton colliders, see, for example, \cite{DKT}--\cite{KL}.  This 
has become more important since LEP is no longer available for Higgs hunting.  An 
obvious advantage of the rapidity gap approach is the spectacularly clean experimental 
signatures (hadron-free zones between the remnants of the incoming protons and the 
produced system) and the possibility to clearly differentiate between different production 
mechanisms.

Events with large rapidity gaps may be selected either by using a calorimeter or by detecting 
leading protons with beam momentum fractions $x_p$ close to 1.  If the momenta of the 
leading protons can be measured with very high precision then a centrally produced state may 
be observed as a peak in the spectrum of the missing-mass ($M$) distribution.  Indeed, it has 
recently been proposed \cite{AR} to supplement CDF with very forward detectors to measure 
both proton and antiproton in Run II of the Tevatron in events with the fractional momentum 
loss $\xi = 1 - x_p$ $\lapproxeq 0.1$ with extremely good accuracy, corresponding to 
missing-mass resolution $\Delta M \simeq 250$~MeV.  The experimental proposal is focused 
on the searches for the Higgs boson, possible manifestations of the physics beyond the 
Standard Model as well as on unique studies of some subtle aspects of QCD dynamics.  It 
is expected that, in association with the high $x_p$ protons and antiprotons, a central system 
with mass up to about 200~GeV can be produced.  To perform such challenging measurements the outgoing 
$p$ and $\bar{p}$ must be detected 
in roman pots, or microstations as proposed for an extension of ATLAS \cite{N}.

To ascertain whether a Higgs signal can be seen, it is crucial to evaluate the 
background.  Recall that the inclusive search for an intermediate mass Higgs, that is $pp$ or 
$p\bar{p} \rightarrow HX$ with $H \rightarrow b\bar{b}$, has an extremely small 
signal-to-background ratio, which makes this process impossible to observe.  If we specialize 
to double-diffractive $H \rightarrow b\bar{b}$ production,  the background is suppressed \cite{KMR3} 
(see also Section~3).

\section{Double-diffractive hard production processes}

Here we are going to present the estimates of the cross-sections for high energy processes of 
the type
\be
\label{eq:a1}
pp \; \rightarrow \; p \: + \: M \: + \: p,
\ee
and similarly for $p\bar{p}$, where a \lq plus\rq\ signs indicates the presence of a large 
rapidity gap.  To be precise, we calculate the rate for the double-diffractive exclusive 
production of a system of large invariant mass $M$, say a Higgs boson.  Our discussion 
below will be focused on the case of an intermediate mass Higgs which dominantly decays 
into the $b\bar{b}$ final state.  From the outset we would like to make it clear that 
today there is no consensus within the community regarding the evaluation of the 
double-diffractive production cross sections, see \cite{PVL,MMB}.  The literature shows a 
wide range of predictions varying by many orders of magnitude\footnote{The range of values 
for exclusive process (\ref{eq:a1}) can be found in 
\cite{KMR1,KMR2,KL,AR,KMR3,MMB}.}.  As in 
Refs.~\cite{BL,KMR1,KMR2,EML,KMR3,KMR4}, we adopt the two-gluon exchange 
picture of the Pomeron, where the amplitude for the double-diffractive process is shown in 
Fig.~1.  Here the hard subprocess $gg \rightarrow M$ is initiated by gluon-gluon fusion and 
an additional relatively soft $t$-channel gluon is needed to screen the colour flow across the rapidity gap 
intervals.

\begin{figure}[t]
\begin{center}
\epsfig{figure=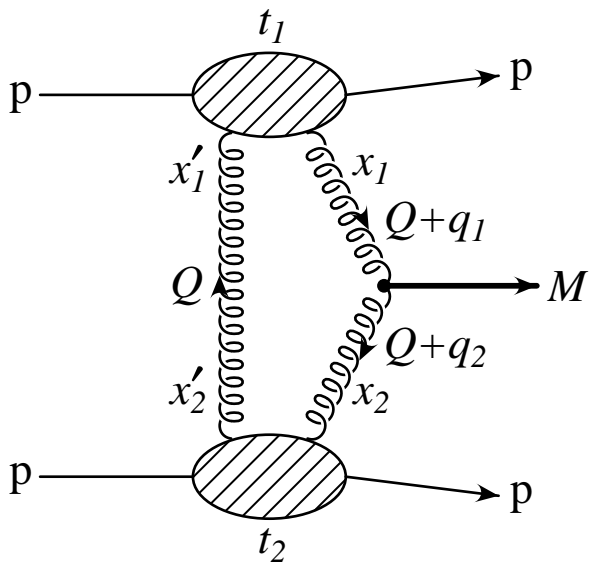}
\end{center}
\caption{
Schematic diagram of double-diffractive production of a system of 
invariant mass $M$, that is the process $pp \rightarrow p + M + p$.}
\end{figure}

One major difference between the various theoretical approaches concerns the specification 
of the exchanged gluons.  {\it Either} non-perturbative gluons are used in which the 
propagator is modified so as to reproduce the total cross section \cite{BL,EML}, {\it or} a 
perturbative QCD estimate is made \cite{KMR2} using an unintegrated, skewed gluon 
density that is determined from the conventional gluon obtained in global parton analyses.  
In this respect, it has been emphasized \cite{AB} (see also \cite{KMR1,KMR2} that the 
non-perturbative normalisation based on the value of the elastic or total cross section fixes the 
diagonal gluon density at $\hat{x} \sim \ell_T/\sqrt{s}$ where the transverse momentum 
$\ell_T$ is small, namely $\ell_T < 1$~GeV.  Thus the value of $\hat{x}$ is even smaller 
than
\be
\label{eq:a2}
x^\prime \; \approx \; \frac{Q_T}{\sqrt{s}} \; \ll \; x \; \approx \; \frac{M}{\sqrt{s}},
\ee
where the variables are defined in Fig.~1.  However, the gluon density grows as $x 
\rightarrow 0$ and so the use of a non-perturbative normalisation will lead to an 
overestimation of double-diffractive cross sections.

Of course the fusion of the two energetic gluons into a high mass state, 
as shown in Fig.~1, is generally accompanied by the emission of soft gluons which may populate 
the rapidity gaps.  The basic mechanism which suppresses this effect is the additional $t$-channel 
gluon of Fig.~1.  It screens the colour, but does not couple to the produced state of mass $M$.   
It has typical values of $Q_T$ which are much smaller than $M$, but which are large enough (for 
sufficiently large $M$) to screen soft gluon emission and to justify the applicability of perturbative QCD.

Another crucial numerical difference between the approaches concerns the size of the 
so-called survival probability of rapidity gaps, $W$, which symbolically can be written as
\be
\label{eq:a3}
W \; = \; S^2 T^2.
\ee
$S^2$ is the probability that the gaps are not filled by secondary particles generated by soft 
rescattering, see for instance \cite{DKS,B,FS,KMR1,KMR2,MMB,GLM,KMR5,KMR6}.  The second factor, 
$T^2$, is the price to pay for not having gluon radiation in the hard subprocess $gg \rightarrow M$.  
The estimates of the survival factor $S^2$ strongly depend on the models for soft hadron rescattering, 
which causes the main uncertainty in evaluation of the double-diffractive cross section.  In \cite{KMR2} 
for $M_H = 120$~GeV at the LHC we tabulated our results for an optimistic estimate $S^2 = 0.1$, while our 
detailed recent calculations \cite{KMR5} yield a lower value, $S^2 = 0.02$.  However it has been pointed 
out \cite{KMR6,KMR2} that it is possible to check the value of $S^2$ by observing 
double-diffractive dijet production\footnote{A promising idea of probing the gap survival factor in the 
$WW$-fusion events with a rapidity gap on either proton side was advocated in Ref.~\cite{CZ}.  For this 
purpose it was proposed there to measure the $Z^0$-production rate with the rapidity gap signature.}.  
This process is driven by the same dynamics, and has a higher cross section, so a comparison of the 
measurements with the predictions can determine $S^2$.  The recent CDF data for diffractive dijet 
production, reported at this Workshop \cite{SNOW,CDF}, appear to be consistent with our determination 
of $S^2$ in Ref.~\cite{KMR6}.  This is clear evidence in favour of the strong suppression due to the 
low survival probability of the rapidity gaps.

The $p\porpbar \rightarrow p + H + \porpbar$ cross section, corresponding to the basic mechanism shown 
in Fig.~1, recently has been calculated to single $\log$ accuracy \cite{KMR2}.  The amplitude is
\be
\label{eq:a4}
{\cal M} \; = \; A \pi^3 \: \int \: \frac{d^2 Q_T}{Q_T^4} \: f_g (x_1, x_1^\prime, Q_T^2, M_H^2/4) 
\: f_g (x_2, x_2^\prime, Q_T^2, M_H^2/4),
\ee
where the $gg \rightarrow H$ vertex factor $A^2$ is given by (\ref{eq:a9}) below, and the unintegrated 
skewed gluon densities are related to the conventional distributions by
\be
\label{eq:a5}
f_g (x, x^\prime, Q_T^2, M_H^2/4) \; = \; R_g \: \frac{\partial}{\partial \ln Q_T^2} \: 
\left [ \sqrt{T (Q_T, M_H/2)} \: xg (x, Q_T^2) \right ].
\ee
The factor $R_g$ is the ratio of the skewed $x^\prime \ll x$ integrated gluon distribution 
to the conventional one \cite{SHUV}.  $R_g \simeq 1.2 (1.4)$ at LHC (Tevatron) energies.  
The bremsstrahlung survival probability $T^2$ is given by
\be
\label{eq:a6}
T (Q_T, \mu) \; = \; \exp \left ( - \: \int_{Q_T^2}^{\mu^2} \: \frac{dk_T^2}{k_T^2} \: 
\frac{\alpha_S (k_T^2)}{2 \pi} \: \int_0^{1 - k_T/\mu} \: dz \: \left [z \: P_{gg} (z) \: 
+ \: \sum_q \: P_{qg}(z) \right ] \right ).
\ee
Note, that the factor $\sqrt{T}$ arises in (\ref{eq:a5}) because the survival probability is 
only relevant to the hard gluon exchanges in Fig.~1.  The origin of the factor $1/Q_T^4$ in the 
integrand in (\ref{eq:a4}) reflects the fact that the production is mediated by the \lq fusion\rq\ 
of two colourless dipoles of size $d^2 \sim 1/Q_T^2$.  Due to the presence of this factor it is 
argued (see e.g.~\cite{KL}) that a perturbative treatment of process (\ref{eq:a1}) is inappropriate.  
However, it is just the Sudakov suppression factor $T$ in the integrand, which makes the integration 
infrared stable and hence the perturbative predictions reliable\footnote{Moreover, the effective anomalous 
dimension, $\gamma$, of the gluon distribution $(xg (x, Q_T^2) \sim (Q_T^2)^\gamma)$ additionally suppresses 
the contribution from the low $Q_T^2$ domain \cite{KMR1}.}.  The saddle points of the integral are located 
near $Q_T^2 = 3.2 (1.5)$~GeV$^2$ at LHC (Tevatron) energies.

The bremsstrahlung survival factor $T$ determines the probability {\it not} to emit the \linebreak 
bremsstrahlung gluons in the interval $Q_T \lapproxeq k_T \lapproxeq M_H/2$.  The upper bound of 
$k_T$ is clear, and the lower bound occurs because there is destructive interference of the amplitude 
in which the bremsstrahlung gluon is emitted from a \lq\lq hard\rq\rq\ gluon with that in which it is 
emitted from the screening gluon.  That is there is no emission when $\lambda \simeq 1/k_T$ is larger 
than the separation $d \sim 1/Q_T$ of the two $t$-channel gluons in the transverse plane, since then 
they act effectively as a colour-singlet system.  The neglect of such an important dampening factor 
in practically all theoretical papers on the double-diffractive Higgs or dijet 
production\footnote{The only exceptions are our results in Refs.~\cite{KMR1,KMR2,KMR3} and the 
evaluation presented in \cite{EML,KL}.  However there is a clear difference in the estimate of the 
survival factor $T^2$ even between these two groups.  The calculation in \cite{KMR1,KMR2,KMR3} yields 
a significantly lower value of $T^2$ than that advocated in \cite{EML,KL}.} reminds one of the attempt 
to force two camels to go through the eye of a needle.  The size of the eye is given by $1/M$, while the 
camel's height is regulated by the position of the saddle point, $1/Q_T$.

 The amplitude (\ref{eq:a4}) corresponds to the exclusive process (\ref{eq:a1}).  The modification for 
 the inclusive process 
\be
\label{eq:a7}
pp \; \rightarrow \; X \: + \: M \: + \: Y
\ee
is given in \cite{KMR1,KMR2,KMR4}, where it was found that the event rate is much larger.  However, 
in the inclusive case the large multiplicity of secondaries poses an additional problem in identifying 
the Higgs boson.

\section{Signal-to-background ratio for double-diffractive Higgs production}

In order to use the \lq missing-mass\rq\ method to search for an intermediate mass Higgs boson, via the 
$H \rightarrow b\bar{b}$ decay mode, we have to estimate the QCD background which arises from the production 
of a pair of jets with invariant mass about $M_H$, see Ref.~\cite{KMR3} for details.  The good news is that 
the signal-to-background ratio does not depend on the uncertainty in the soft survival factor $S^2$, and is 
given just by the ratio of the corresponding $gg \rightarrow H \rightarrow b\bar{b}$ and $gg \rightarrow 
b\bar{b}$ subprocesses.  

We begin by assuming that the $b$ jets are not tagged.  Then the main background is the double-diffractive 
colour-singlet production of a pair of high $E_T$ gluons with rapidities $\eta_1$ and $\eta_2$.  The 
$gg \rightarrow gg$ subprocess cross section is \cite{KMR4,BC}
\be
\label{eq:a8}
\frac{d \hat{\sigma}}{d^2 p_T} \; = \; \frac{9 \alpha_S^2}{8 p_T^6} \: \left ( \frac{M^4}{4 p_T^4} 
\: - \: \frac{M^2}{p_T^2} \right )^{- \frac{1}{2}} \; \frac{dM^2}{d (\Delta \eta)},
\ee
where $M$ is the invariant mass of the dijet system, $p_T$ is the transverse momentum of the jets, 
and $\Delta \eta = | \eta_1 - \eta_2 |$ is the jet rapidity difference.  The background (\ref{eq:a8}) 
should be compared to the double-diffractive $gg \rightarrow H$ signal
\be
\label{eq:a9}
\frac{A^2}{4} \; = \; \frac{\sqrt{2}}{36 \pi^2} \; G_F \: \alpha_S^2.
\ee
First of all, in order to reduce the background we impose a jet $E_T$-cut.  For instance, if we trigger 
on events containing a pair of jets with angles $\theta > 60^\circ$ from the proton direction in the 
Higgs rest frame, then we only eliminate one half of the signal, whereas the background dijet cross 
section is reduced to
\be
\label{eq:a10}
\frac{d \hat{\sigma}}{dM^2} \; = \; 9.7 \: \frac{9 \alpha_S^2}{8 M^4}.
\ee
With a common scale for the coupling $\alpha_S$, and neglecting the NLO corrections, we obtain 
a signal-to-background ratio
\be
\label{eq:a11}
\frac{S}{B_{gg}} \; = \; (4.3 \times 10^{-3}) \: {\rm Br} (H \rightarrow b\bar{b}) \; 
\left ( \frac{M}{100~{\rm GeV}} \right )^3 \; \left ( \frac{250~{\rm MeV}}{\Delta M} \right ).
\ee
If $M_H = 120$~GeV, the ratio $S/B_{gg} \sim 5 \times 10^{-3}$.  This is too small for the above 
approach to provide a viable signal for the Higgs boson.  However the situation is greatly improved 
if we are able to identify the $b$ and $\bar{b}$ jets.  If we assume that there is only a 1\% chance 
to misidentify a gluon jet as a $b$ jet, then tagging {\it both} the $b$ and $\bar{b}$ jets will suppress 
the gluon background by $10^4$.  In this case only the true $b\bar{b}$ background may pose a problem.

A remarkable advantage of the double-rapidity-gap signature for the $H \rightarrow b\bar{b}$ events 
is that here the $H \rightarrow b\bar{b}$ signal/$b\bar{b}$ background ratio is strongly enhanced due 
to the colour factors, gluon polarization selection and the spin $\frac{1}{2}$ nature of the quarks 
\cite{KMR3}.  First, the background $b\bar{b}$-dijet rate is suppressed due to the absence of the 
colour-octet $b\bar{b}$-state.  Thus, for $E_T^2 < M^2/4$ we have
\be
\label{eq:a12}
\frac{d \hat{\sigma} (gg \rightarrow b\bar{b})}{d \hat{\sigma} (gg \rightarrow gg)} \; < \; 
\frac{1}{4 \times 27} \; < \; 10^{-2}.
\ee
Second, we emphasize that for the exclusive process the initial $gg$ state obeys special 
selection rules.  Besides being a colour-singlet, for forward outgoing protons the projection of the 
total angular momentum is $J_z = 0$ along the beam axis.  On the other hand, the Born amplitude for 
light fermion pair production\footnote{For light quark pair exclusive production $p + p \rightarrow 
p + q\bar{q} + p$, with forward outgoing protons, the cancellation was first observed by Pumplin 
\cite{PUMPLIN}, see also \cite{BC,KMR4}.} vanishes in this $J_z = 0$ state, see, for example, 
\cite{INOK}.  This result follows from $P$- and $T$-invariance and fermion helicity conservation 
of the $J_z = 0$ amplitude \cite{BKSO}.  Thus, if we were to neglect the $b$-quark mass $m_b$, 
then at leading order we would have no QCD $b\bar{b}$-dijet background at all.  Even beyond LO, 
the interference between the signal and background amplitudes is negligibly small, since they have 
different helicity structure.  Therefore the form of the peak, observed in 
double-diffractive exclusive $H \rightarrow b\bar{b}$ production, will not be affected by interference 
with $b\bar{b}$ jets produced by the pure QCD background process.

Of course, a non-vanishing $b\bar{b}$ background is predicted when we allow for the quark mass or if 
we emit an extra gluon.  Nevertheless in the former case we still have an additional suppression to 
(\ref{eq:a12}) of about a factor of $m_b^2/p_T^2 \simeq 4 m_b^2/M_H^2 < 10^{-2}$, whereas in the 
latter case the extra suppression is about $\alpha_S/\pi \simeq 0.05$.  Note that events containing 
the third (gluon) jet may be experimentally separated from Higgs decay, where the two jets are dominantly 
co-planar\footnote{The situation here is similar to the signal-to-background ratio for intermediate mass
 Higgs production in polarised $\gamma\gamma$ collisions, which was studied in detail in \cite{BKSO,MSK}.}.  
 However, the price to pay for this separation is the further reduction of the signal caused by the 
 Sudakov suppression of the final state radiation in the Higgs events \cite{BKSO}.

Thus, the two-gluon fusion mechanism for hard production, illustrated in Fig.~1, provides a 
unique situation where the incoming hard two-gluon system is practically fully polarized with $J_z = 0$. 
 An explicit calculation \cite{KMR3}, assuming $M_H = 120$~GeV and imposing the $\theta > 60^\circ$ cut 
 of low $E_T$ jets, gives a signal-to-background ratio
\be
\label{eq:a13}
\frac{S}{B_{b\bar{b}}} \; \gapproxeq \; 4 \: \left ( \frac{1~{\rm GeV}}{\Delta M} \right ).
\ee
The signal is, thus, in excess of background even at mass resolution $\Delta M \sim 2$~GeV, so 
the $b\bar{b}$ background should not be a problem\footnote{Unfortunately the situation worsens 
for inclusive Higgs production, when the polarization arguments become redundant.  In this 
case $S/B_{b\bar{b}}$ ratio is additionally suppressed by a factor $\sim 20-30$.}.

\section{Is the production rate of Higgs events with rapidity gaps large enough?}

While the predictions for the $S/B_{b\bar{b}}$-ratio look quite favourable for Higgs searching 
using the missing-mass method, the expected event rate casts a shadow on the feasibility of this 
approach (at least for experiments at the Tevatron).  The cross section for exclusive double-diffractive 
Higgs production at Tevatron and LHC energies has been calculated by several authors\footnote{A more 
complete set of references to related theoretical papers can be found in Ref.~\cite{AR}.} 
\cite{BL,KMR1,KMR2,EML,KL}.  In our recent analysis \cite{KMR5} the gap survival probability 
for the double-diffractive process is estimated to be $S^2 = 0.05$ at $\sqrt{s} = 2$~TeV and 
$S^2 = 0.02$ at $\sqrt{s} = 14$~TeV.  If we incorporate these estimates into the perturbative 
QCD calculations of \cite{KMR2}, we find
\bea
\label{eq:a14}
\sigma_H & = & \sigma (p\bar{p} \rightarrow p + H + \bar{p}) \; \simeq \; 0.06~{\rm fb} 
\quad\quad {\rm at} \quad\quad \sqrt{s} \; = \; 2~{\rm TeV}, \\
\label{eq:a15}
\sigma_H & = & \sigma (pp \rightarrow p + H + p) \; \simeq \; 2.2~{\rm fb} \quad\quad {\rm at} 
\quad\quad \sqrt{s} \; = \; 14~{\rm TeV}
\eea
for a Higgs boson of mass 120~GeV.  These values are much lower than the expectations of other 
authors listed in \cite{AR}.  However, as we already mentioned, the recent CDF study of diffractive
 dijet production \cite{SNOW,CDF}, provides strong experimental evidence in favour of our pessimistic 
 estimates of the survival factor $S^2$, see \cite{KMR6}.  CDF \cite{SNOW,CDF} have studied 
 double-diffractive dijet production for jets with $E_T > 7$~GeV.  They find an upper limit for 
 the cross section, $\sigma({\rm dijet}) < 3.7$~nb, as compared to our prediction of about 
 1~nb \cite{KMR6}.  Using the dijet process as a monitor thus rules out the much larger predictions 
 for $\sigma (p\bar{p} \rightarrow p + H + \bar{p})$ which exist in the literature.  Unfortunately 
 the prediction $\sigma_H \simeq 0.06$~fb of (\ref{eq:a14}) means that Run II of the Tevatron, with 
 an integrated luminosity of ${\cal L} = 15$~fb$^{-1}$, should yield less than an event.  We should 
 add that the double-diffractive Higgs search can also be made in the $\tau^+ \tau^-$ and $WW^*$ 
 decay channels, but, due to the small branching ratios, the event rate is even less.

We emphasize that such a low expected signal cross section at the Tevatron just illustrates the 
high price to be paid for improving the $S/B_{b\bar{b}}$ ratio by selecting events with double 
rapidity gaps.  On the other hand, a specific prediction of the perturbative approach is that 
the cross section $\sigma_H$ steeply grows with energy \cite{KMR1,KMR2} (c.f.\ (\ref{eq:a14}) 
with (\ref{eq:a15})), in contrast to non-perturbative phenomenological models based on Ref.~\cite{BL}.  
In fact, if we were to ignore the rapidity gap survival probability, $S^2$, then $\sigma_H$ would 
have increased by more than a factor of 100 in going from $\sqrt{s} = 2$~TeV to $\sqrt{s} = 14$~TeV.  
However, at the larger energy, the probability to produce secondaries which populate the gaps 
increases, and as a result the $\sigma (pp \rightarrow p + H + p)$ increases only by a factor 
of 40.  Nevertheless, there is a real chance to observe double-diffractive Higgs production at 
the LHC, since both the cross section and the luminosity are much larger than at the Tevatron.  
Another test of our perturbative scenario is the behaviour of the dijet cross section with the 
jet $E_T$.  Due to the $x$ dependence of the perturbative gluon, we predict a steeper fall off 
with increasing $E_T$ than the 
non-perturbative models \cite{KMR2}.

For the inclusive production of a Higgs of mass $M_H = 120$~GeV we expect, at the LHC energy, 
a cross section of the order of 40(4)~fb, taking rapidity gaps $\Delta \eta = 2(3)$ \cite{KMR2}.

The double-diffractive dijet cross sections are much larger than those for Higgs production.  
For example, if we take a dijet bin of size $\delta E_T = 10$~GeV for each jet and $\eta_1 = 
\eta_2$ we obtain, for $E_T = 50$~GeV jets at LHC energies,
\be
\label{eq:a16}
\left . d \sigma_{\rm excl}/d\eta \right |_0 \; \simeq \; 40~{\rm pb}, \quad \left .
 d \sigma_{\rm incl}/d\eta \right |_0 \; \simeq \; 250~{\rm pb},
\ee
where $\eta \equiv (\eta_1 + \eta_2)/2$.  The rapidity gaps are taken to be $\Delta 
\eta$(veto) $= (\eta_{\rm min}, \eta_{\rm max}) = (2,4)$ for the inclusive case (see 
\cite{KMR4} for the definition of the kinematics).  Such a high event rate and the 
remarkable purity of the di-gluon system, that is generated in the exclusive double-diffractive 
production process, provides a unique environment to make a detailed examination of high energy 
gluon jets\footnote{In double-diffractive exclusive high-$E_T$ dijet events the jets appear to 
be pure gluon ones at the level about 3000:1.  Moreover, after an appropriate selection of the 
two-jet configuration and the removal of the $b\bar{b}$ contamination by tagging, the sample 
may become (at least) an order of magnitude purer.}.  Indeed, we may speak here of a \lq gluon 
factory\rq\ \cite{KMR3}.

Let us finally comment on the soft suppression factor $S^2$, which causes the main uncertainty 
in various calculations of the rate of rapidity gap events and, thus, is the \lq Achilles heel\rq\ 
of the evaluation of the Higgs production signal in the exclusive and inclusive processes of 
(\ref{eq:a1}) and (\ref{eq:a7}) respectively.  This factor depends sensitively on the spatial 
distributions of partons inside the proton, and thus, is closely related to the whole diffractive 
part of the $S$-matrix, see Ref.~\cite{KKMR} for details.  For example, the survival probability 
for central Higgs production by $WW$ fusion, with large rapidity gaps on either side, is expected 
to be larger than that for the double-Pomeron exchange \cite{KMR2,KKMR}.  Recall that a quantitative 
probe of the suppression factor $S^2$ can be achieved either by measurements of central $Z$ production 
\cite{CZ} or of dijet production.

An instructive example is Higgs boson production by the $\gamma\gamma \rightarrow H$ fusion 
subprocess \cite{KP}.  This process takes place at very large impact parameters, where the 
corresponding gap survival probabilities are $S^2 = 1$ \cite{KMR2,KMR7}.  $\sigma (\gamma\gamma 
\rightarrow H)$ is estimated to be about 0.03~fb at $\sqrt{s} = 2$~TeV and 0.3~fb at $\sqrt{s} = 
14$~TeV, which is comparable to our expectations (\ref{eq:a14}), (\ref{eq:a15}) for double-diffractive 
Higgs production.  Note that the strong and electromagnetic contributions have negligible interference, 
because they occur at quite different values of the impact parameter.

As well known \cite{EP}, the two-photon mechanism of Higgs production is especially interesting 
for heavy ion collisions where the photon flux scales with $Z^2$ ($Z$ being the charge of the 
nucleus) \cite{LL}.  In the photon-photon case for $M_H = 120$~GeV we expect a signal-to-background 
ratio $S/B_{b\bar{b}} \sim 1$, if the experimental resolution on the reconstruction of the 
$b\bar{b}$-invariant mass is $\sim 10$~GeV, and if a cut $| \cos \theta | < 0.7$ is imposed on the 
$b$ and $\bar{b}$ jets.

\section{Conclusions}

We have examined the possibility of performing a high resolution missing-mass search for the 
Higgs boson at the Tevatron, that is the process $p\bar{p} \rightarrow p + H + \bar{p}$ where 
a \lq plus\rq\ denotes a large rapidity gap.  Using a model-independent approach, we find that 
there is a huge QCD background arising from double-diffractive dijet production.  A central detector 
to trigger on large $E_T$ jets is essential.  Even so, the signal-to-background ratio is too small
 for a viable \lq missing-mass\rq\ Higgs search.  The situation is much improved if we identify the 
 $b$ and $\bar{b}$ jets.  The $gg \rightarrow H \rightarrow b\bar{b}$ signal is now in excess of the 
 QCD $gg \rightarrow b\bar{b}$ background, even for a mass resolution of $\Delta M \sim 2$~GeV.  
 The only problem is that, when proper account is taken of the survival probability of the rapidity 
 gaps, the $p\bar{p} \rightarrow p + H + \bar{p}$ event rate is too small at the Tevatron.  Recall 
 that the experimental limit on the cross section for double-diffractive dijet production confirms 
 the small predicted rates.  Nevertheless, there is a real chance to observe double-diffractive Higgs
  production at the LHC, since both the cross section and the luminosity are much larger than at the Tevatron.

The rather pessimistic expectations of the missing-mass Higgs search at the Tevatron are, however, 
compensated by a by-product of the double-diffractive proposal.  The 
double-diffractive production of dijets offers a unique {\it gluon factory}, generating 
huge numbers of essentially pure gluon jets from a colour-singlet state in an exceptionally clean environment.

\section*{Acknowledgements}

We thank M. Albrow, D. Goulianos, A.B. Kaidalov, R. Orava, A. Rostovtsev and W.J. Stirling 
for useful discussions.  VAK thanks The Leverhulme Trust for a Fellowship.  This work was also 
supported by PPARC, the Russian Fund for Fundamental Research 
(98-01-17629) and the EU Framework TMR programme, contract FMRX-CT98-0194 
(DG-12-MIHT).


\newpage
\begin{thebibliography}{xx}
\bibitem{DKT} Yu.L. Dokshitzer, V.A. Khoze and S.I. Troyan, Sov. J. Nucl. Phys. {\bf 46} (1987) 712.
\bibitem{DKS} Yu.L. Dokshitzer, V.A. Khoze and T. Sj\"{o}strand, Phys. Lett. {\bf B274} (1992) 116.
\bibitem{B} J.D. Bjorken, Int. J. Mod. Phys. {\bf A7} (1992) 4189; Phys. Rev. {\bf D47} (1993) 101.
\bibitem{BL} A. Bialas and P.V. Landshoff, Phys. Lett. {\bf B256} (1991) 540.
\bibitem{FS} R.S. Fletcher and T. Stelzer, Phys. Rev. {\bf D48} (1993) 5162 and references therein.
\bibitem{KMR1} V.A. Khoze, A.D. Martin and M.G. Ryskin, Phys. Lett. {\bf B401} (1997) 330.
\bibitem{KMR2} V.A. Khoze, A.D. Martin and M.G. Ryskin, Eur. Phys. J. {\bf C14} (2000) 525.
\bibitem{EML} E.M. Levin, {\tt hep-ph/9912403} and references therein.
\bibitem{KL} D. Kharzeev and E.M. Levin, {\tt hep-ph/0005311}.
\bibitem{AR} M.G. Albrow and A. Rostovtsev, {\tt hep-ph/0009336}.
\bibitem{N} V. Nomokonov, these proceedings.
\bibitem{KMR3} V.A. Khoze, A.D. Martin and M.G. Ryskin, {\tt hep-ph/0011393} and references therein.
\bibitem{PVL} P.V. Landshoff, these proceedings.
\bibitem{MMB} M.M. Block, these proceedings.
\bibitem{KMR4} V.A. Khoze, A.D. Martin and M.G. Ryskin, Phys. Rev. {\bf D56} (1997) 5867.
\bibitem{AB} A. Berera, Phys. Rev. {\bf D62} (2000) 014015.
\bibitem{GLM} E. Gotsman, E.M. Levin and U. Maor, Phys. Lett. {\bf B353} (1995) 526; Phys. Rev. 
{\bf D60} (1999) 094011 and references therein.
\bibitem{KMR5} V.A. Khoze, A.D. Martin and M.G. Ryskin, Eur. Phys. J. {\bf C18} (2000) 167.
\bibitem{KMR6} V.A. Khoze, A.D. Martin and M.G. Ryskin, {\tt hep-ph/0006005}; 
{\tt hep-ph/0007083}, Phys. Lett. {\bf B} (in press).
\bibitem{CZ} H. Chehime and D. Zeppenfeld, Phys. Rev. {\bf D47} (1993) 3898.
\bibitem{SNOW} G. Snow, these proceedings.
\bibitem{CDF} CDF Collaboration:  T. Affolder et al., Phys. Rev. Lett. {\bf 85} (2000) 4215.
\bibitem{SHUV} A.G. Shuvaev, K.J. Golec-Biernat, A.D. Martin and M.G. Ryskin, Phys. Rev. {\bf D60} (1999) 014015.
\bibitem{BC} A. Berera and J.C. Collins, Nucl. Phys. {\bf B474} (1996) 183.
\bibitem{PUMPLIN} J. Pumplin, Phys. Rev. {\bf D52} (1995) 1477.
\bibitem{INOK} K.A. Ispiryan, I.A. Nagorskaya, A.G. Oganesyan and V.A. Khoze, Sov. J. Nucl. Phys. {\bf 11} (1970) 712.
\bibitem{BKSO} D.L. Borden, V.A. Khoze, W.J. Stirling and J. Ohnemus, Phys. Rev. {\bf D50} (1994) 4499; \\
V.S. Fadin, V.A. Khoze and A.D. Martin, Phys. Rev. {\bf D56} (1997) 484.
\bibitem{MSK} M. Melles, W.J. Stirling and V.A. Khoze, Phys. Rev. {\bf D61} (2000) 054015.
\bibitem{KKMR} A.B. Kaidalov, V.A. Khoze, A.D. Martin and M.G. Ryskin, to be published.
\bibitem{KP} K. Piotrzkowski, these proceedings.
\bibitem{KMR7} V.A. Khoze, A.D. Martin, R. Orava and M.G. Ryskin, 
{\tt hep-ph/0010163}, Eur. Phys. J. {\bf C} (in press).
\bibitem{EP} E. Papageorgiu, Phys. Lett. {\bf B352} (1995) 394.
\bibitem{LL} L.D. Landau and E.M. Lifshitz, Sov. Phys. {\bf 6} (1934) 247.
\end{thebibliography}
\end{document}